\documentclass[aip,
 amsmath,amssymb,
 reprint,%
]{revtex4-1}

\usepackage{graphicx}% Include figure files
\usepackage{dcolumn}% Align table columns on decimal point
\usepackage{bm}% bold math
\usepackage[utf8]{inputenc}
\usepackage[T1]{fontenc}
\usepackage{mathptmx}
\usepackage{etoolbox}
\usepackage{siunitx}
\usepackage{booktabs}

\newcommand{\MarkUp}[1]{\color{black} #1 \color{black}}
\begin{document}

\preprint{AIP/123-QED}

\title{Scattering–Induced Loss in Ferroelectric Photonic Devices}
% Force line breaks with \\
\author{Jonah Townsend}
\affiliation{Department of Physics and Astronomy, University of Victoria, Victoria, British Columbia, Canada V8W 2Y2}
\affiliation{Department of Physics and Astronomy, University of British Columbia, Vancouver, British Columbia, Canada V6T 1Z1}
\author{Enzo Conceição Picinini}%
\affiliation{Department of Physics and Astronomy, University of Victoria, Victoria, British Columbia, Canada V8W 2Y2}
\affiliation{Department of Physics and Astronomy, University of British Columbia, Vancouver, British Columbia, Canada V6T 1Z1}
\author{Rogério de Sousa}
\thanks{Author to whom correspondence should be addressed: \href{mailto:rdesousa@uvic.ca}{rdesousa@uvic.ca}}
\affiliation{Department of Physics and Astronomy, University of Victoria, Victoria, British Columbia, Canada V8W 2Y2}
\affiliation{Centre for Advanced Materials and Related Technology, University of Victoria, Victoria, British Columbia V8W 2Y2, Canada}
%\email{rdesousa@uvic.ca}
\date{\today}% It is always \today, today,

\begin{abstract}
Ferroelectric materials have colossal optical nonlinearities, but their integration into quantum photonic chips is made challenging by the additional loss mechanisms that they introduce.
Here we present a perturbative theory that expresses non-absorptive (elastic) photon scattering–induced loss as a functional of a general spectral density for spatial fluctuations of electric permittivity. We apply the theory to calculations of attenuation coefficients $\alpha$ in slab waveguides in order to compare two distinct loss mechanisms: Interface roughness and ferroelectric domain disorder. 
Our theory can account for realistic roughness without special symmetry considerations, and it demonstrates how to use electron microscopy images of ferroelectric domains to obtain explicit numerical predictions for $\alpha$. 
Loss is maximum when the mean domain length is comparable to the wavelength of light (Mie regime), indicating that, for telecom wavelengths, sub-micron domains (Rayleigh regime) or single domain waveguides provide equivalent strategies for reducing loss. 
\end{abstract}

\maketitle

Quantum computers based on photons are innately compatible with telecommunication networks, enabling communication with high security and collaboration with other quantum computers in a future quantum internet.\cite{Luo2023}
However, developing quantum computers with a large number of qubits is a great challenge,\cite{PSiQuantum2025, Bourassa2021, Simmons2024} requiring improvements to photonic chip technology to achieve large optical nonlinearity for qubit control at the same time as maintaining extremely low loss to preserve quantum information.\cite{Marchant2024, Anderson2025, Ulrich2025, Kim2026}

Recent experiments show that barium \MarkUp{titanate} (BaTiO$_3$ or BTO), the textbook ferroelectric, can be grown epitaxially on silicon photonic chips and possesses nonlinear optics (Pockels) coefficients that are 20 times larger than lithium niobate (LiNbO$_3$ or LNO), the commonly used ferroelectric in photonic devices.\cite{Abel2019, Chelladurai} 
However, a major difference between BTO and LNO is that the former cannot be made into a single ferroelectric domain with uniform polarization $\bm{P}$ across the waveguide.\cite{Chelladurai} As a result, the electrical permittivity changes in space, breaking translation symmetry and leading to elastic photon scattering. This causes loss of power from the dominant guided mode into other unguided and guided modes.\cite{Jackson}

\begin{figure}
    \centering
\includegraphics[width=1\linewidth]{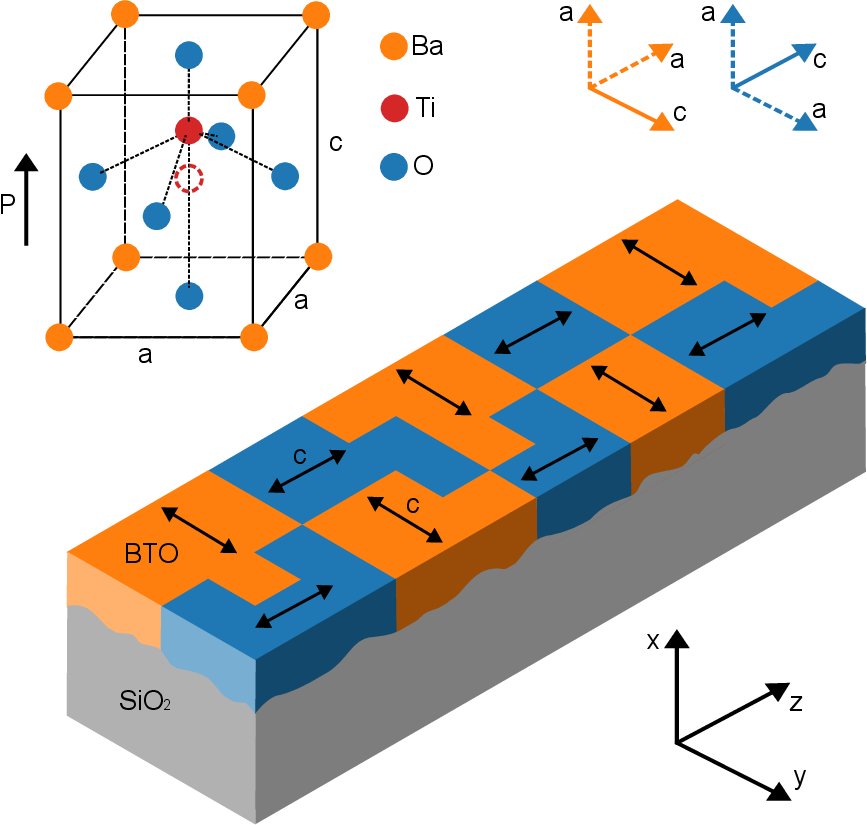}
    \caption{Diagram of a barium titanate (BTO) on silicon dioxide (SiO$_2$) dielectric waveguide, where z is the direction of propagation, depicting interface roughness and stochastic distribution of domains with orthogonal optical axes. The BTO unit cell displayed in the top left corner shows the a- and c-axes, as well as the polarization $P$ as a result of the displaced titanium ion.}
    \label{fig:Diagram}
\end{figure}

In a waveguide experiencing photon loss, the intensity of light along the propagation direction $z$ is given by $I(z) = I_0e^{-\alpha z}$, where $\alpha$ is called the attenuation coefficient. Here we derive an expression for the attenuation coefficient associated with two types of random fluctuations in ferroelectric waveguides: Interface roughness and domain disorder (Fig.~\ref{fig:Diagram}).

Loss due to roughness of dielectric interfaces has been extensively studied, and several models for the attenuation coefficient have been developed.\cite{Marcuse, PayneLacey, Hormann} In this work, we propose a model for roughness loss that is agnostic to the waveguide geometry and roughness autocorrelation function, generalizing previous theories that focused on one-dimensional exponential correlations for roughness.\cite{Hormann}

In contrast, the loss caused by ferroelectric domains is an emerging theme in ferroelectric photonics. A recent theory\cite{Kim2025b} concluded that domain walls dominate non-absorptive loss, and predicted $> 10^{-4}$~dB loss \emph{per domain wall} in BTO devices. With typical average domain size $\bar{L}\sim \qty{10}{\nm}$,\cite{Ju2025, Reynaud2025} this would imply an attenuation coefficient greater than $10^{2}$~dB/cm, making the BTO platform too lossy for quantum applications. 
This result is contradicted by measurements in BTO waveguides that reported total loss of the order of $1$~dB/cm, $100\times$ lower.\cite{Riedhauser2025, Kim2025a}

In this letter, we propose a perturbative electrodynamical method that expresses elastic loss as a functional of the spectral density for spatial variations of the permittivity inside the core waveguide material.
We use electron microscopy images of the domain distribution in BTO\cite{Ju2025, Reynaud2025} to estimate the spectral density, obtaining explicit predictions for the attenuation coefficient as a function of wavelength. The theory shows that nonabsorptive loss emerges as a result of average domain size length scales ($\bar{L}$), with the nanometre length scales of domain walls playing a minor role. 

%{\bf Attenuation coefficient from perturbation theory}.--

\MarkUp{Our first goal is to obtain the attenuation coefficient from perturbation theory. To do this, we write} the sourceless Ampère-Maxwell law at fixed frequency $\omega$ to first order with small perturbations in the electric field $\bm E$, magnetic field $\bm H$, and permittivity $\epsilon$, 
\begin{equation}
    \label{eq:Perturbation_Ampere_Maxwell}
    {\bm\nabla} \times \delta{\bm H} = -i\omega\delta\epsilon{\bm E}^{(0)} - i\omega\epsilon^{(0)}\delta{\bm E}.
\end{equation}
The term $-i\omega\delta\epsilon{\bm E}^{(0)}$ functions as an effective volume current, generating the fields $\delta{\bm E}$ and $\delta{\bm H}$. Assuming the unperturbed modal field ${\bm E}^{(0)}$ has harmonic $z$-dependence with propagation constant $\beta$, 
we calculate the 
%far-field 
vector potential due to the effective current source in the free-space approximation for the unguided modes\cite{Jackson, Snyder_and_Love}
\begin{align}
{\bm A}({\bm r}) = &  -\frac{i\mu_0\omega}{4\pi}\frac{e^{ik_\text{cl}r}}{r}\int d^3r'e^{-ik_\text{cl}{\bm \hat{r}\cdot r'}}e^{i\beta z'}{\bm E}^{(0)}({\bm r}'_t)\delta\epsilon({\bm r}'), \label{free_space_approx}
\end{align}
where $k_\text{cl}$ is the cladding wavenumber, $\bm{\hat{r}}=\bm{r}/r$, and 
$\bm{r}_t=(x,y)$ 
denotes directions transverse to the propagation direction $z$ (Fig.~\ref{fig:Diagram}). 
\MarkUp{Equation~(\ref{free_space_approx}) shows that in the presence of $\delta\epsilon$, the guided mode $\bm{E}^{(0)}$ generates fields that decay as $1/r$ in the far field outside the waveguide.
Thus, it describes photon loss to unguided modes \emph{only}, although it can be generalized to account for scattering out to other guided modes by using a Green's function with near fields}. 

From now on we focus on photon loss to unguided modes; we plan to address loss to other guided modes in future work. The associated Poynting vector is calculated by the average radiated fields of an ensemble of waveguides with stochastic permittivity variations $\delta \epsilon(y,z)$ and fixed waveguide length $\ell$.
We integrate the Poynting vector over the surface of a sphere far away from the waveguide in all directions, yielding the average radiated power. Finally, after dividing by $\ell$ 
and taking the limit $\ell \to \infty$, we arrive at
\begin{eqnarray}
\alpha_{\delta\epsilon} &=&  \frac{\mu_0 n_\text{cl}\omega^{4}}{64\pi^3cP_{\rm guide}}\int d\Omega_r \int_{-\infty}^{\infty}dq_y\nonumber\\
&&\times\left|\tilde{\bm{E}}^{(0)}\left(q_y\hat{y}+k_\text{cl}\hat{\bm{r}}\cdot\hat{\bm{x}}\hat{\bm{x}}\right)\times\bm{\hat{r}}\right|^{2}\nonumber\\
&&\times \tilde{C}_{\delta\epsilon}
\left[k_\text{cl}\left(\hat{\bm{r}}-\hat{\bm{r}}\cdot\hat{\bm{x}}\hat{\bm{x}}\right)-q_y\hat{y}-\beta\hat{z}\right]\label{eq:alpha_bulk},
\end{eqnarray}
where $d\Omega_r$ is the differential solid angle, $n_\text{cl}$ is the cladding refractive index, $\tilde{\bm E}^{(0)}$ is the Fourier transform of the unperturbed modal electric field inside the waveguide core, 
$P_\text{guide}=\frac12 \int d^2r_t ( {\bm E}^{(0)} \times {{\bm H}^{(0)}}^* ) \cdot {\bm \hat z}$ is the initial guided power, and  
\begin{equation}
    \tilde{C}_{\delta\epsilon}(\bm{q}_\|)=\int d^2r_{\|} e^{-i\bm{q}_\|\cdot(\bm{r}_{\|}-\bm{r}'_{\|})}\langle \delta\epsilon(\bm{r}_{\|})\delta\epsilon(\bm{r}'_{\|})\rangle
    \label{2dcorrfnc}
\end{equation}
is the spectral density of permittivity spatial fluctuations over the waveguide interface $\bm{r}_{\|}=(y,z)=r_{\|}(\sin{\theta},\cos{\theta})$.

\begin{figure*}
    \includegraphics[width=\linewidth]{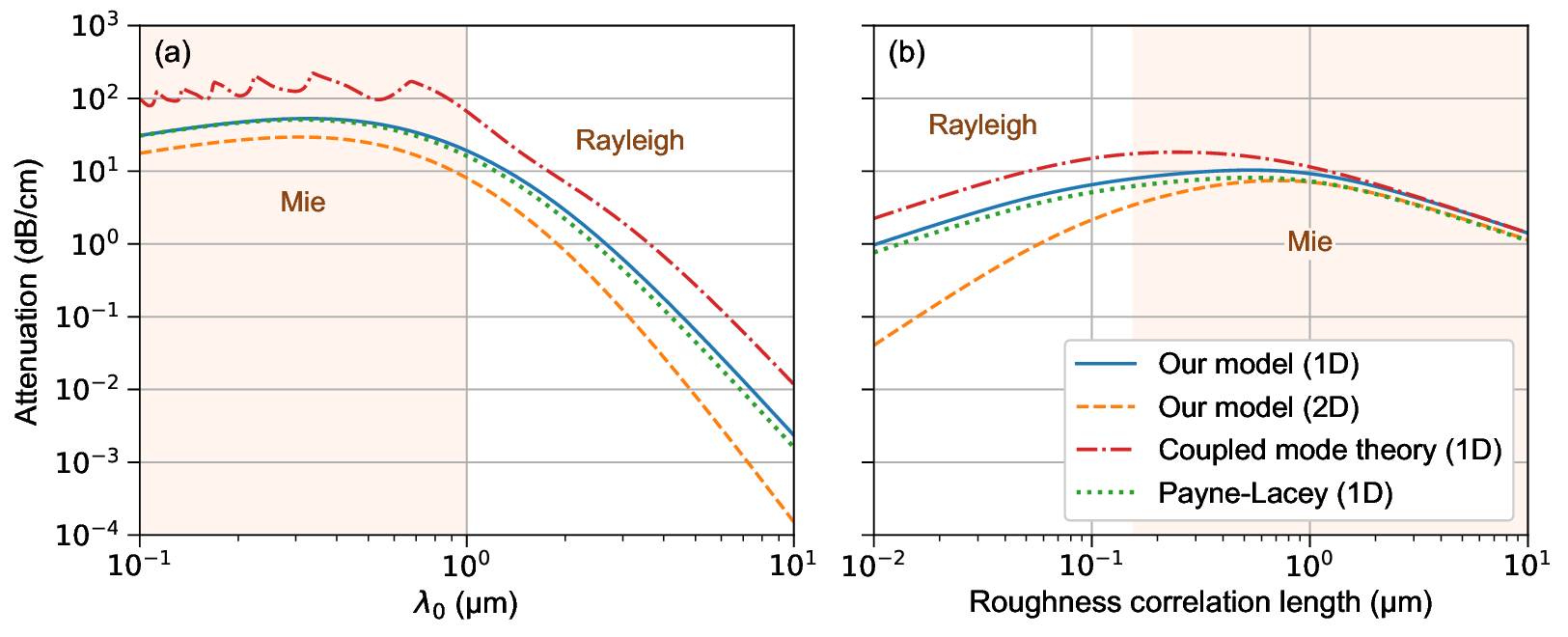}
\caption{\label{fig:RoughnessLoss} Attenuation $\alpha_{\delta h}$ for the lowest TE mode\cite{Snyder_and_Love} of a BTO planar slab waveguide due to roughness on the bottom BTO/SiO$_2$ interface. (a) $\alpha_{\delta h}$ vs. vacuum wavelength $\lambda_0$ for roughness correlation $L_\text{c}=\qty{0.1}{\micro\metre}$ and standard deviation $\sigma_h= \qty{2}{\nano\metre}$.\cite{Roberts} (b) $\alpha_{\delta h}$ vs. $L_\text{c}$ for $\lambda_0=\qty{1.55}{\micro\metre}$ and $\sigma_h= \qty{2}{\nano\metre}$. Other parameters shown in Table~\ref{TableParameters}. We compare the 1D roughness model (height fluctuations $\delta h\equiv \delta h(z)$, with roughness along propagation direction $z$ \emph{only}) to the more realistic 2D roughness model with $\delta h\equiv\delta h(y,z)$. For 1D, we show that our model falls right in-between the coupled mode theory\cite{Marcuse} and Payne-Lacey\cite{PayneLacey} models. In contrast, our more realistic 2D model leads to attenuation that is systematically smaller than all 1D models.}
\end{figure*}

For roughness of a single waveguide interface, we relate perturbations in the height $\delta h$ and permittivity $\delta\epsilon$ to first order via $\delta\epsilon = \frac{d\epsilon^{(0)}}{dh}\delta h({\bm r}_\|)$.\cite{Hormann}
Following the approach of Johnson et al.,\cite{Johnson} we smooth the permittivity and take the limit as the smoothing parameter $s$ goes to zero, taking $\bm \hat x$ as the surface normal, giving
$f\left[\bar{\epsilon}(x)\right] = \int_{-\infty}^\infty dx' f\left[\epsilon(x')\right]g_s(x - x')$,
where $g_s$ is any smoothing function that goes to a Dirac delta as the smoothing parameter $s$ goes to zero.
We do this smoothing anisotropically, such that the electric field components parallel to the waveguide interface see an arithmetically smoothed permittivity, with $f(\epsilon) = \epsilon$, whereas the perpendicular component sees ``logarithmic'' smoothing, with $f(\epsilon) = \ln(\epsilon)$. This allows us to write $\delta\epsilon{\bm E}^{(0)}$ in terms of only the field components which are well-defined at the interface:
\begin{align}
    \delta\epsilon{\bm E}^{(0)} = & \left[ \ln\left(\frac{\epsilon_\text{co}}{\epsilon_\text{cl}}\right)\bm{D}_\perp^{(0)} + (\epsilon_\text{co} - \epsilon_\text{cl})\bm{E}_\|^{(0)} \right]\nonumber\\ 
    &\times \delta h({\bm r}_\|) \delta(x - d)\nonumber \\
    \equiv &\, {\bm G}(y)\delta h({\bm r}_\|)\delta(x - d),\label{source_surface}
\end{align}
where $\epsilon_\text{co}$ and $\epsilon_\text{cl}$ are the permittivities in the core and cladding respectively, and the rough interface is located at $x=d$.

The derivation then proceeds similarly to that for $\delta\epsilon$ fluctuations, except that the radiated field is now in the $yz$ plane, propagating along $k_\text{cl}\bm{\hat{r}}_{\|}=k_\text{cl}\bm{r}_{\|}/r_{\|}$. For a single rough interface, we get
\begin{eqnarray}
    \alpha_{\delta h} &=& \frac{\mu_0n_\text{cl}\omega^4}{64\pi^3cP_\text{guide}} \int_{0}^{2\pi} d\theta\int_{-\infty}^\infty dq_y \left|\tilde{\bm G}(q_y) \times {\bm \hat r}\right|^2\nonumber\\
    &&\times \tilde{C}_{\delta h}\left[k_\text{cl}\bm{\hat{r}}_{\parallel} - q_y{\bm \hat y} - \beta{\bm \hat z}\right], \label{eq:alpha_rough}
\end{eqnarray}
where $\tilde{C}_{\delta h}(\bm{q}_\|)=\int d^2 r_{\|} e^{-i\bm{q}_\|\cdot(\bm{r}_{\|}-\bm{r}'_{\|})}\langle \delta h(\bm{r}_{\|})\delta h(\bm{r}'_{\|})\rangle$ is the spectral density for interface roughness. 
To get the total attenuation caused by multiple rough interfaces, we can simply add their individual attenuation coefficients; since their roughnesses are uncorrelated, the radiation produced by them is incoherent.

Equations~(\ref{eq:alpha_bulk})~and~(\ref{eq:alpha_rough}) generalize previous theories\cite{Marcuse, PayneLacey} to arbitrary input guided modes and dielectric/height spectral densities. 
They can be used with numerically simulated modal fields to calculate loss in any waveguide geometry.
Below we present explicit numerical calculations for infinite planar slab waveguides for which the guided modes are known analytically, allowing direct comparison to previous theories. Note that when considering a planar slab waveguide with lateral width $W$ and length $\ell$ both taken to be $\infty$, Eqs.~(\ref{eq:alpha_bulk}),~(\ref{eq:alpha_rough}) assume implicitly that the emitted radiation occurred at a sphere of radius $R$ that is taken to $\infty$ \emph{before} $W$ and $\ell$ are taken to $\infty$. The order of limits ensures losses to unguided modes radiating in all directions are accounted for. In that sense our results for infinitely planar slabs are a proxy for rectangular waveguides with finite width $W\gg \lambda_0$.

Neglecting the near fields in Eq.~(\ref{free_space_approx}) has the advantage of 
simpler interpretation in terms of elastic scattering of guided photons. For Eq.~(\ref{eq:alpha_bulk}), the guided photon has wavevector 
$\bm{k}_{\rm guide} =q_y\bm{\hat{y}} + k_\text{cl}\bm{\hat{r}}\cdot\bm{\hat{x}}\bm{\hat{x}}+\beta\bm{\hat{z}}$ (argument of $\tilde{\bm{E}}^{(0)}$ plus $\beta\bm{\hat{z}}$), while the photon radiated into the cladding has ${\bm k}_\text{rad} = k_\text{cl}\bm{\hat{r}}$. The change in momentum that a guided photon experiences when scattering into the cladding is then $\Delta {\bm k} = {\bm k}_\text{rad} - {\bm k}_\text{guide} = k_\text{cl}\left(\bm{\hat{r}}-\bm{\hat{r}}\cdot\bm{\hat{x}}\bm{\hat{x}}\right) - q_y\bm{\hat{y}} - \beta\bm{\hat{z}}$. This difference is exactly the argument of the spectral density that appears in Eq.~(\ref{eq:alpha_bulk}), indicating that the permittivity fluctuations provide the change in momentum needed to scatter a photon from a guided to an unguided mode.
Note how the $x$ component of $\bm{k}_{{\rm guide}}$ is unchanged during scattering.
Equation~(\ref{eq:alpha_rough}) has a similar interpretation, with the difference that only guided photons localized at $x=d$ possessing zero momentum along $x$ contribute to scattering, $\bm{k}_{{\rm guide}}=q_y\bm{\hat{y}}+\beta\bm{\hat{z}}$. This happens because of $\delta(x-d)$ in the radiation source Eq.~(\ref{source_surface}), which leads to $\bm{k}_{{\rm rad}}=k_\text{cl}\bm{\hat{r}}_{\parallel}$. 

\begin{table}
%\centering
\caption{\textbf{BTO waveguide parameters (Figs.~\ref{fig:RoughnessLoss}~and~\ref{fig:BulkLoss}}).}
\label{TableParameters}
\begin{tabular}{lc}
\toprule
Slab half width $d$ & $\qty{0.1}{\um}$\\
Core index $n_\text{co}$ & $\sqrt{(n_{o}^{2} + n_{e}^{2})/2}$\\
Ordinary index $n_0=\sqrt{\epsilon_a/\epsilon_0}$ (BTO)\cite{Abel2019} & $2.301$\\
Extraordinary index $n_\text{e}=\sqrt{\epsilon_c/\epsilon_0}$ (BTO)\cite{Abel2019} & $2.271$\\
Cladding index $n_\text{cl}$ (SiO$_2$)\cite{Haynes} & $1.55$\\
%\begin{tabular}{@{}l@{}}Propagation constant $\beta$\cite{Snyder_and_Love} \\ (from implicit eqn with $k_j=n_j 2\pi/\lambda_0$)\end{tabular}
Implicit Eq. for prop. const. $\beta$\cite{Snyder_and_Love}
& \hspace{-7ex}$\sqrt{\frac{\beta^{2}-k_\text{cl}^{2}}{k_\text{co}^{2}-\beta^{2}}}=\tan{\left(d\sqrt{k_\text{co}^{2}-\beta^{2}}\right)}$\\
\bottomrule
\end{tabular}
\end{table}

%{\bf Loss due to interface roughness}.-- 

\MarkUp{We now present explicit numerical results for loss due to interface roughness.} 
Figure~\ref{fig:RoughnessLoss} shows calculations of Eq.~(\ref{eq:alpha_rough}) for the ``1D model'' autocorrelation $\langle \delta h(\bm{r}_{\|})\delta h(\bm{r}'_{\|})\rangle = \sigma^{2}_{h} e^{-|z-z'|/L_\text{c}}$ for one rough interface, \MarkUp{using parameters from Table~\ref{TableParameters}}. It shows that our theory falls right in between coupled-mode\cite{Marcuse} and Payne-Lacey\cite{PayneLacey} theories, demonstrating consistency with previous approximations. 

Coupled mode theory goes beyond the free space approximation of Eq.~(\ref{free_space_approx}) by including the near field of unguided modes.\cite{Marcuse, Snyder_and_Love} This leads to oscillations in the Mie regime of scattering ($\lambda_0\lesssim 10 L_\text{c}$ in Fig.~\ref{fig:RoughnessLoss}(a)), and also makes $\alpha$ larger by a factor of $\sim 2$. 

It is more realistic to assume the ``2D model'' $\langle \delta h(\bm{r}_{\|})\delta h(\bm{r}_{\|}')\rangle = \sigma^{2}_{h} e^{-|\bm{r}_{\|}-\bm{r}_{\|}'|/L_\text{c}}$.
As shown in Fig.~\ref{fig:RoughnessLoss}, this 2D model gives rise to attenuation that is consistently lower than the 1D model. 
For the 1D model, $\tilde{C}_{\delta h}(k_\text{cl}\bm{\hat{r}}_{\|}- q_y{\bm \hat y} - \beta{\bm \hat z})=2\pi\delta(k_\text{cl}\sin{\theta}-q_y)\tilde{C}_{\delta h}^{1D}(k_{cl}\cos{\theta}-\beta)$, 
which has a Lorentzian resonance at $\beta=k_\text{cl}\cos{\theta}$. For the slab waveguide only $q_y=0$ contributes, and the $\delta(k_\text{cl}\sin{\theta})$ implies only forward ($\theta=0$) and back ($\theta=\pi$) scattering is allowed. Thus $\alpha_{\delta h}$ for the 1D model is dominated by a forward scattering resonance at $\beta=k_\text{cl}$. 
In contrast, the 2D model's roughness in the $y$ direction provides momentum transfer along $k_y$, enabling scattering along all angles $\theta$. This smooths out the Lorentzian resonance at $\theta=0$ leading to a reduction of $\alpha_{\delta h}$. 

\begin{figure*}
\includegraphics[width=\linewidth]{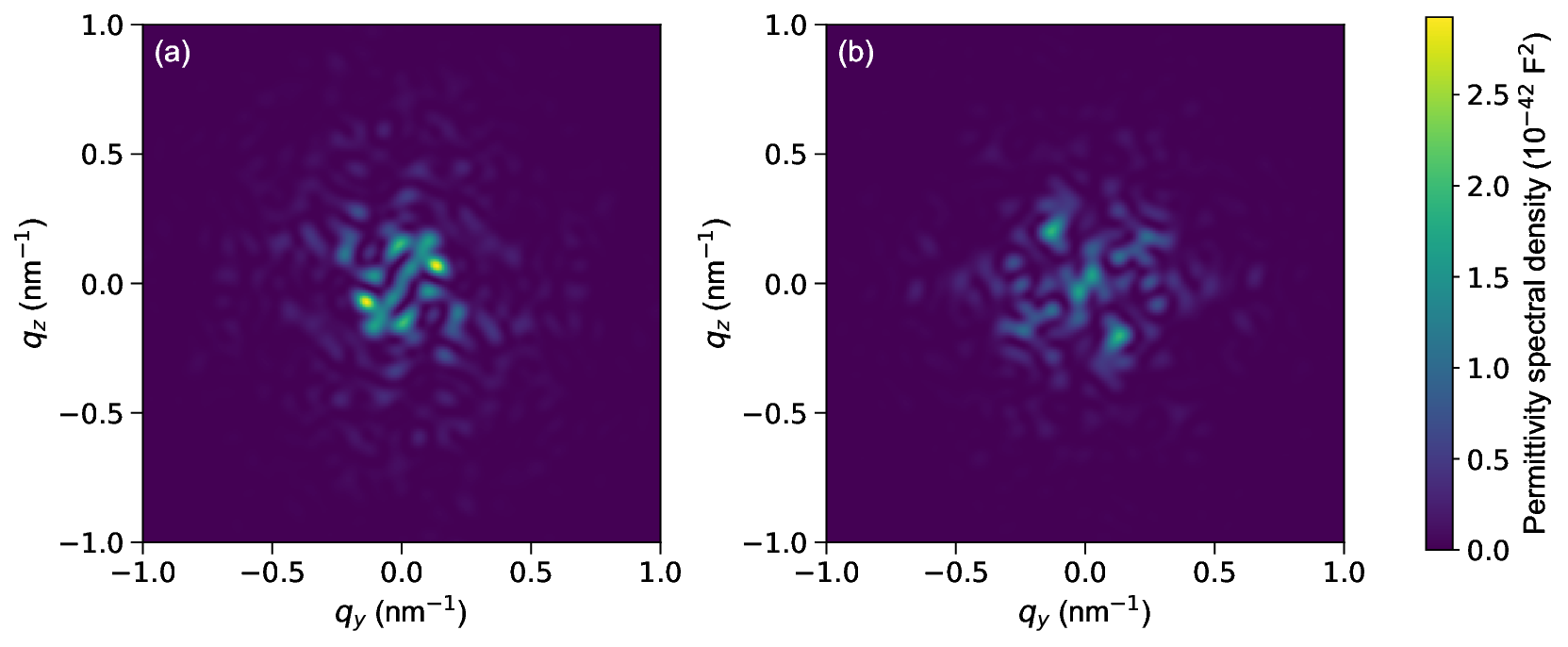}
\caption{Spectral density $\tilde{C}_{\delta\epsilon}(\bm{q}_\|)$ for dielectric constant fluctuations due to ferroelectric domains in BTO, for (a) slow cooled and (b) fast cooled growth conditions.\cite{Ju2025} The spectral densities were obtained by mapping electron microscopy data (Fig.~1c in \onlinecite{Ju2025}) to $\epsilon$ and taking the Fourier transform. Each figure shows two large primary peaks at $\pm \bm{Q}_0$ that demonstrate nearly stripe order for the domains.\label{Fig2dSD}}
\end{figure*}

%{\bf Loss due to ferroelectric domain disorder}.-- 

\MarkUp{We now turn to explicit numerical predictions of loss due to ferroelectric domains.}
In order to evaluate loss using Eq.~(\ref{eq:alpha_bulk}), we need to obtain an expression for the spectral density $\tilde{C}_{\delta\epsilon}(\bm{q}_\|)$ associated with the ferroelectric domain structure. The actual domain configuration in ferroelectric photonic devices depends on growth conditions and is a topic of current research. 
In Ref.~\onlinecite{Ju2025}, electron microscopy images of lattice parameter ratios $\xi$ for slow and fast cooling rates during film growth over a $0.2 \times \qty{0.2}{\,(\um)^{2}}$ area (Fig.~1c in \onlinecite{Ju2025}) reveal a mix of $a_y$ ($\xi=c/a>1$) and $a_z$ ($\xi=a/c<1$) domains, \MarkUp{along with nm-scale domain walls/islands where $\xi$ is measured to be continuously varying between $c/a$ and $a/c$.
%along with their domain walls. 
Our Fig.~\ref{fig:Diagram} depicts the $a_y$ and $a_z$ domains in blue and orange respectively, with domain walls not shown.} 
%
%The correspond to the blue and orange domains in our Fig.~\ref{fig:Diagram}. 
A TE mode (electric field $\bm{E}\| \hat{\bm{y}}$) will experience $\epsilon=\epsilon_a$ and $\epsilon=\epsilon_c$ when propagating in the blue ($a_y$) and orange ($a_z$) domains, respectively. \MarkUp{These lead to propagation with ordinary index $n_o=\sqrt{\epsilon_a/\epsilon_0}$ in the $a_y$, and extraordinary index $n_e=\sqrt{\epsilon_c/\epsilon_0}$ in the $a_z$}.

\MarkUp{Here we propose a linear mapping to connect $\xi$ to $\epsilon$, 
\begin{equation}
\epsilon(\xi) = [(\xi-a/c)\epsilon_a+(c/a-\xi)\epsilon_c]/(c/a-a/c).
\label{eq:mapping}
\end{equation}
Such a mapping implies $|\delta\epsilon|/\epsilon_0<(\epsilon_a-\epsilon_c)/\epsilon_0=0.1372$ in the domain boundary region $a/c<\xi<c/a$. The fact that the latter value is quite similar to predictions from density functional theory ($\delta\epsilon/\epsilon_0=0.11$)\cite{Kim2025b} suggests our linear mapping remains a good approximation even inside domain boundaries.} 
%
%Given that $(\epsilon_a-\epsilon_c)/\epsilon_0=0.1372\ll 1$,\cite{Abel2019} it is reasonable to assume a linear mapping to connect $\xi$ to $\epsilon$: $\epsilon(\xi) = [(\xi-a/c)\epsilon_a+(c/a-\xi)\epsilon_c]/(c/a-a/c)$. 
%
Applying this linear mapping and taking the Fourier transform\cite{NoteSD} leads to the estimated spectral densities $\tilde{C}_{\delta\epsilon}(\bm{q})$ shown in our Figs.~\ref{Fig2dSD}(a) (slow cooled) and \ref{Fig2dSD}(b) (fast cooled). 

The spectral densities show primary peaks at $\pm \bm{Q}_0$, on top of secondary peaks that are much lower in magnitude and extend to higher harmonics. We interpret the primary peaks as describing a domain distribution with quasi-stripe order; in contrast, the secondary peaks extend to higher $\bm{q}$ and are interpreted as small-length-scale features such as domain walls \MarkUp{and islands}. 

\begin{figure*}
\includegraphics[width=\linewidth]{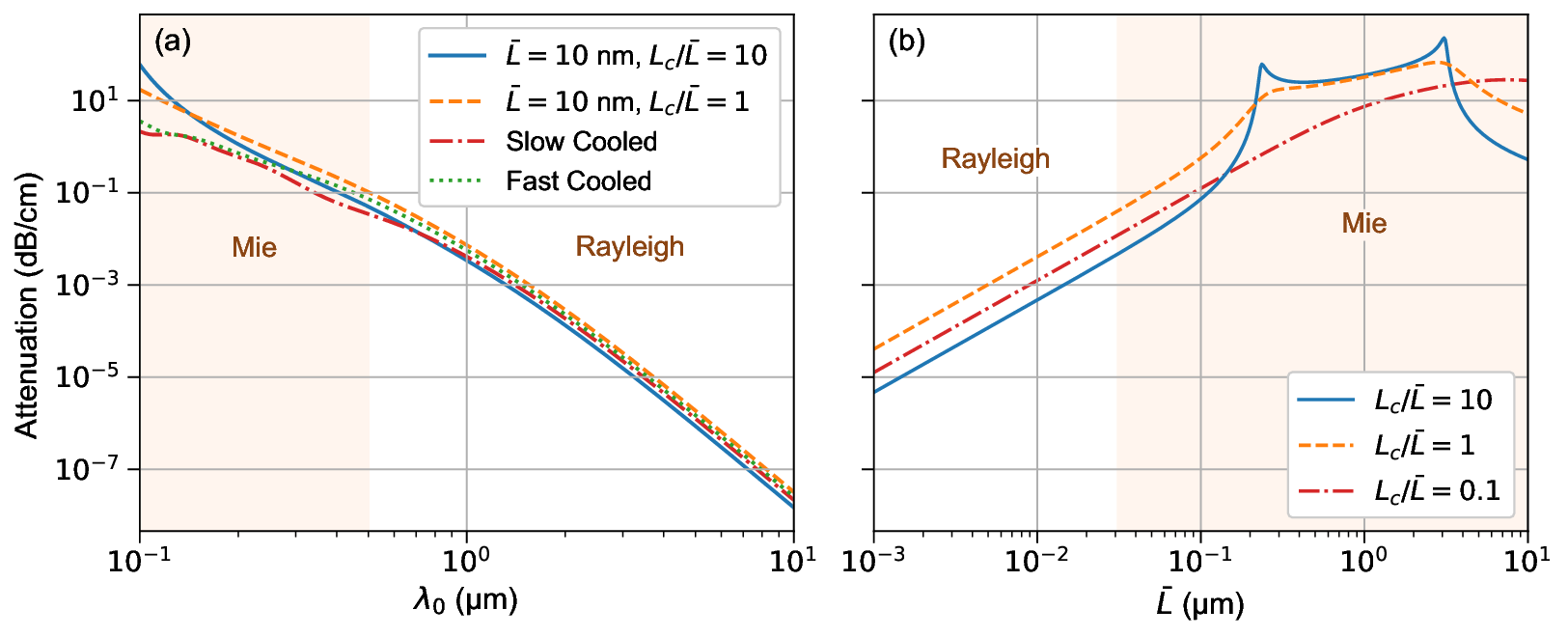}
\caption{\label{fig:BulkLoss} Attenuation $\alpha_{\delta\epsilon}$ in a single-mode BTO planar waveguide due to ferroelectric domains, using the fields of the lowest order TE guided mode.\cite{Snyder_and_Love} 
(a) Calculated $\alpha_{\delta\epsilon}$ vs. $\lambda_0$ using Eq.~(\ref{eq:alpha_bulk}). The slow and fast cooled $\alpha_{\delta\epsilon}$ assumed spectral density $\tilde{C}_{\delta\epsilon}(\bm{q}_\|)$ directly interpolated from the values of Fig.~\ref{Fig2dSD}(a,b). \MarkUp{These are seen to be in good agreement with the simple model Eq.~(\ref{2dLorentzianModel}) 
that neglects the domain wall peaks and assumes average domain size $\bar{L}=10$~nm and domain correlation length $L_\text{c}=100$~nm}.
(b) $\alpha_{\delta\epsilon}$ vs. $\bar{L}$ for $\lambda_0=\qty{1.55}{\micro\metre}$ (telecom wavelength) and different $L_\text{c}/\bar{L}$. \MarkUp{The attenuation is seen to have a broad maximum in the Mie regime $\bar{L}>\lambda_0/50$, due to the presence of two resonances that arise from our assumption of domain disorder restricted to the $yz$ plane (see text)}.
Assumed parameters in Table~\ref{TableParameters}.}
\end{figure*}

Figure~\ref{fig:BulkLoss}(a) shows explicit calculations of the attenuation coefficient as a function of $\lambda_0$ using Eq.~(\ref{eq:alpha_bulk}) \MarkUp{and the parameters of Table~\ref{TableParameters}, with the spectral density obtained from direct numerical interpolation of Figs.~\ref{Fig2dSD}(a,b)}. 
%
% along with numerical interpolation of the spectral densities of
%
We find that a simpler model that includes only the two primary peaks is able to reproduce these results quite well:
\begin{eqnarray}
    \tilde{C}_{\delta\epsilon}(\bm{q}_\|)&=&\left(\frac{\epsilon_a-\epsilon_c}{2}\right)^{2} \nonumber\\
   &&\hspace{-5em} \times\left\{\frac{\pi\Gamma}{\left[\left(\bm{q}_\|-\bm{Q}_0\right)^{2}+\Gamma^{2}\right]^{3/2}}
    +\frac{\pi\Gamma}{\left[\left(\bm{q}_\|+\bm{Q}_0\right)^{2}+\Gamma^{2}\right]^{3/2}}\right\},
\label{2dLorentzianModel}
\end{eqnarray}
where $\bm{q}_{\|}=(q_y,q_z)$, $Q_0=\pi/\bar{L}$ with $\bar{L}$ the average domain size, and $\Gamma=1/L_\text{c}$ with $L_\text{c}$ the correlation length for the domains. In real space such a model corresponds to $C_{\delta\epsilon}(\bm{r}_{\|})\propto \cos{\left(\bm{Q}_0\cdot\bm{r}_{\|}\right)}e^{-|r_{\|}|/L_\text{c}}$. Figure~\ref{fig:BulkLoss}(a) shows Eq.~(\ref{2dLorentzianModel}) with $\bar{L}=10$~nm and $L_\text{c}=100$~nm fits the data quite well in a wide wavelength range. \MarkUp{The ability of the primary peaks [Eq.~(\ref{2dLorentzianModel})] to fit our predicted $\alpha_{\delta\epsilon}$ obtained from numerical interpolation of the electron microscopy data gives quantitative evidence that domain walls/islands play a negligible role in non-absorptive scattering when $\lambda_0>50 \bar{L}$, the Rayleigh regime in Fig.~\ref{fig:BulkLoss}(a)}.

%This shows that the secondary peaks interpreted as coming from domain walls play a negligible role. 

Inspection of Eq.~(\ref{eq:alpha_bulk}) reveals why this is the case: The small length scale of a domain wall implies large values of $\bm{q}_\|$ in $\tilde{C}_{\delta\epsilon}(\bm{q}_\|)$; but for these large values to satisfy photon momentum \MarkUp{and energy conservation} during scattering, the guided mode $\tilde{\bm{E}}^{(0)}(\bm{k}_{{\rm guide}})$ must supply an equivalently large $\bm{k}_{{\rm guide}}$, which is suppressed for smooth core modes.

Figure~\ref{fig:BulkLoss}(b) presents the attenuation as a function of $\bar{L}$ for different values of $L_\text{c}/\bar{L}$, for telecom $\lambda_0=\qty{1.55}{\um}$. The attenuation is seen to have a broad maximum across a band of values \MarkUp{in the Mie regime $\bar{L}>\lambda_0/50$}. This again can be understood from the change in momentum supplied by $\tilde{C}_{\delta\epsilon}(\Delta \bm{k})$. In the limit $L_\text{c}\rightarrow \infty$, $\tilde{C}_{\delta\epsilon}(\Delta \bm{k})$ is sharply peaked at $\Delta\bm{k}=\pm \bm{Q}_0$ (the dominant domain pattern), and the angular integration in Eq.~(\ref{eq:alpha_bulk}) leads to a $1/\left|\cos{\phi}\right|$ singularity that appears when we integrate a sharply peaked function that depends on $\sin{\phi}$.  The presence of this singularity implies scattering is \emph{resonant} when $\bm{k}_{{\rm rad}}$ is in the $yz$ plane ($\phi=\pm \pi/2$). Therefore, while all scattering directions are allowed by momentum conservation, $\bm{k}_{{\rm rad}}=k_\text{cl}\bm{\hat{r}}=\bm{k}_{{\rm guide}}+\Delta\bm{k}$, only incident photons with $\bm{k}_{{\rm guide}}\cdot\bm{\hat{x}}=0$ are resonant. Note how this follows from our assumption of domain disorder in the yz plane, restricting $\Delta\bm{k}\cdot\bm{\hat{x}}=0$. Combining the modulus of the momentum conservation with $\bm{k}_{{\rm guide}}\cdot\bm{\hat{x}}=0$ yields the criterion for resonance: $k_\text{cl}=\left|q_y \bm{\hat{y}}+\beta\bm{\hat{z}}\pm \bm{Q}_0\right|$.

The two peaks in Fig.~\ref{fig:BulkLoss}(b) for $L_\text{c}/\bar{L}=10$ can be explained by this resonance. Our calculation assumed $\bm{Q}_0=Q_0\bm{\hat{z}}$, and the slab waveguide contains only $q_y=0$, with $\beta=\qty{7.59}{\um^{-1}}$ and $k_\text{cl}=2\pi n_\text{cl}/\lambda_0=\qty{6.28}{\um^{-1}}$. The criterion for resonance then becomes $Q_0=\beta\pm k_\text{cl}$, leading to $\bar{L}_{+}=\pi/(\beta+k_\text{cl})=\qty{0.22}{\um}$ and $\bar{L}_{-}=\pi/(\beta-k_\text{cl})=\qty{2.40}{\um}$. These values are quite close to the ones seen in Fig.~~\ref{fig:BulkLoss}(b). The fact that $L_\text{c}<\infty$ converts the two peaks into a ``resonance band'' with $\alpha_{\delta\epsilon}$ enhanced in the range $\pi/(\beta+k_\text{cl})\lesssim \bar{L}\lesssim \pi/(\beta-k_\text{cl})$.

When the domain configuration has $\bar{L}\ll \pi/(\beta+k_\text{cl})\approx \lambda_0/[2(n_{{\rm co}}+n_{{\rm cl}})]$, resonant scattering is strongly suppressed, placing $\alpha_{\delta\epsilon}$ in the Rayleigh regime where it is much lower. 

%{\bf Conclusions}.-- 

\MarkUp{In conclusion}, we derived expressions for the attenuation coefficient associated with photon loss to unguided modes due to interface roughness and domain disorder in ferroelectric waveguides in terms of arbitrary unperturbed modal fields and a general spectral density for spatial dielectric fluctuations. 

Numerical calculations of the attenuation due to interface roughness showed that our theory is in good agreement with previous models, which relied on symmetry along $y$, limiting their applicability to planar devices with 1D roughness along $z$.\cite{Marcuse, PayneLacey} The latter is a particularly unphysical assumption, since it implies roughness has infinite correlation ($L_\text{c}=\infty$) along $y$. 

In contrast, our model applies to arbitrary waveguide geometries, modal fields, and 2D roughness autocorrelation along both interface directions $y$ and $z$. Numerical evaluations with the physical 2D roughness showed that attenuation is always smaller than in the 1D case, deviating significantly from it when $L_\text{c}\ll \lambda_0$. 

We also proposed a theory for elastic photon loss due to domain disorder in ferroelectric waveguides, and showed that the relevant spectral density for dielectric fluctuations can be \MarkUp{obtained} from electron microscopy images. The theory shows that loss due to domains can be explained by a simple model that depends on the average domain size $\bar{L}$ and its correlation length $L_\text{c}$, \MarkUp{with nanometer structures such as domain walls and islands giving a negligible contribution to scattering because their large photon momentum ``kicks'' can not satisfy energy and momentum conservation for smooth core modes. The insensitivity to spatial features much smaller than $\lambda_0$ is a general result for non-absorptive scattering out to unguided modes; in particular it applies to any inhomogeneity of $\delta\epsilon$, not just the $a_y$, $a_z$ domains found in BTO}. 

Figure~\ref{fig:BulkLoss}(a) presents explicit numerical predictions for $\alpha_{\delta\epsilon}$ as a function of $\lambda_0$ for the BTO samples probed by electron microscopy:\cite{Ju2025, Reynaud2025} For $\lambda_0=\qty{1.55}{\um}$ we predict $\alpha_{\delta\epsilon}<10^{-3}$~dB/cm. This prediction is consistent with measurements of $1$~dB/cm total loss in BTO waveguides,\cite{Riedhauser2025, Kim2025a} indicating that the measured loss is dominated by roughness coming from the surface, interface, or sidewalls.

Our predicted $\alpha_{\delta\epsilon}$ is several orders of magnitude lower than a recent theory that relied on the finite-difference time domain method (FDTD) to compute the loss due to a single domain wall.\cite{Kim2025b} \MarkUp{We speculate that their overestimation of loss is due to poor convergence, since performing accurate FDTD would require a simulation spanning several micrometers with a mesh fine enough to resolve the nanometer scale domain wall. The large cell count means that a significant number of iterations are needed to reach convergence.} In contrast, our analytical method expresses attenuation as an integral over the spectral density of dielectric fluctuations, accounting for all length scales without numerical convergence challenges. 

% \MarkUp{Their overestimation of loss due to each domain wall might have happened due to convergence issues, because performing accurate FDTD would require a mesh that resolves the domain wall nanometer scale while the simulation spans several micrometers. Convergence requires a significant number of iterations because the required cell count scales badly.}

%We believe the latter theory overestimated the loss due to each domain wall due to its 
%need to use a mesh that resolves nm-scale domain walls while the simulation must span several micrometres, leading to convergence issues. 

The attenuation due to domain disorder is predicted to be resonantly enhanced in the Mie regime when the average domain size $\bar{L}\sim \lambda_0$. 
However, it is instead strongly suppressed when 
\MarkUp{$\bar{L}<\lambda_0/50$}, 
%\bar{L}\ll \lambda_{0}/[2(n_{{\rm co}}+n_{{\rm cl}})]$, 
the Rayleigh regime in Fig.~\ref{fig:BulkLoss}(b). This indicates that, in the design and construction of high-quality ferroelectric photonic devices, it is important to consider not only surface/interface roughness but also the average ferroelectric domain size. One should ensure that domain lengths are outside the range of wavelength of light $\lambda_0$, either by using sub-micron domains or single-domain waveguides. 
In this regime ferroelectric photonics with smooth interfaces has sufficiently low loss to preserve quantum information, opening up applications for a wide variety of quantum devices.

\begin{acknowledgments}
We acknowledge the support of the Natural Sciences and Engineering Research Council of Canada (NSERC), through the Consortium on Integrated Quantum Photonics with Ferroelectric Materials ALLRP 587352-23, and the USRA program. We thank J. Appleby-Millette, S.V. Grayli, T. Lu, I. Paci, P. Wang, and J. Young for discussions and criticisms. 
\end{acknowledgments}

\subsection*{Author Contributions}
RdS proposed the problem; JT, ECP, and RdS formulated the theory, and JT carried out the numerical calculations. All authors contributed equally to the writing of the paper and interpretation of the results.

\section*{Data Availability Statement}

Data sharing is not applicable to this article as no new data were created or analyzed in this study.

% This command includes ALL papers in the .bib file, regardless of whether they were explicitly cited. So this should not be used
%\nocite{*}

\bibliography{aipsamp}% Produces the bibliography via BibTeX.

\end{document}